\documentstyle[12pt,aps,prd,preprint,graphicx]{revtex}
\begin{document}
\title{Mass-Inflation in Dynamical Gravitational Collapse of a  
Charged Scalar-Field}
\author{Shahar Hod and Tsvi Piran}
\address{The Racah Institute for Physics, The
Hebrew University, Jerusalem 91904, Israel}
\date{\today}
\maketitle

\begin{abstract}
We study the inner-structure of a charged black-hole which is formed
from the gravitational collapse of a self-gravitating 
{\it charged} scalar-field.
Starting with a {\it regular} spacetime, we follow the evolution
through the formation of an apparent horizon, a Cauchy horizon and 
a final central singularity.
We find a null, {\it weak}, mass-inflation singularity along the
Cauchy horizon, which is a precursor of a strong, {\it spacelike} 
singularity along the
$r=0$ hypersurface.
\end{abstract}
\bigskip

The {\it no-hair} theorem, introduced by Wheeler \cite{Wheeler} in the
early 1970s, states that the {\it external}
field of a black-hole relaxes to a Kerr-Newman field characterized solely
by the black-hole's mass, charge and angular-momentum. 
This simple picture describing the exterior of a black-hole is in
dramatic contrast with its {\it interior}.
The singularity theorems of Penrose and Hawking \cite{PenHaw} predicts
the occurrence of inevitable spacetime singularities 
as a result of a gravitational
collapse in which a black-hole forms. 
According to the {\it weak cosmic censorship} conjecture \cite{Penrose1} 
these spacetime singularities are hidden beneath 
the black-hole's event-horizon.
However, these theorems tell us nothing about the nature of these
spacetime singularities. In particular, the {\it final} outcome of a
generic gravitational collapse is still an open question in
general relativity.

Until recently, our physical intuition regarding the nature of these
inner singularities was largely based on the known static or
stationary black-hole solutions: Schwarzschild ({\it spacelike},
strong and unavoidable central singularity), Reissner-Nordstr\"om and
Kerr ({\it timelike}, strong singularity).  Further insight was gained
from the work of Belinsky, Khalatnikov and Lifshitz \cite{BKL} who
found a strong oscillatory spacelike singularity. However, a new and
drastically different picture of these inner black-hole singularities
has emerged in the last few years, according to which the CH inside
charged or spinning black-holes is transformed into a {\it null}, {\it
  weak} singularity \cite{Hiscock,Poisson,Ori1,Ori2,Brady}.  The CH
singularity is weak in the sense that an infalling observer which hits
this null singularity experiences only a finite tidal deformation
\cite{Ori1,Ori2}.  Nevertheless, curvature scalars (namely, the
Newman-Penrose Weyl scalar $\Psi_{2}$) diverge along the CH, a
phenomena known as {\it mass-inflation} \cite{Poisson}.  The physical
mechanism which underlies this CH singularity is actually quite
simple: small perturbations, which are remnants of the gravitational
collapse {\it outside} the collapsing object are gravitationally {\it
  blueshifted} as they propagate in the black-hole's interior parallel
to the CH \cite{Penrose2} (the mass-inflation scenario itself includes
in addition an outgoing radiation flux which irradiates the CH. This
outgoing flux represents a portion of the ingoing radiation which is
scattered {\it inside} the black-hole).

Yet, it should be emphasized that
despite of the remarkable progress in our physical understanding
of the {\it inner}-structure of black-holes, 
the evidence supporting the existence of a  null, weak CH singularity 
is mostly based on
{\it perturbative} analysis. Thus, it is of interest to perform a 
{\it full} non-linear investigation of the inner-structure of black-holes.
The pioneering work of Gnedin and Gnedin \cite{Gnedin} was a first
step in this direction. They have demonstrated the existence of a
central {\it spacelike} singularity deep inside a charged black-hole 
coupled to a (neutral) scalar-field. Much insight was
gained from the numerical work of Brady and Smith \cite{Brady}
who studied the non-linear evolution of a (neutral) scalar-field on a
spherical charged black-hole. These authors established the existence
of a {\it null} mass-inflation singularity along the CH. 
Furthermore, they have shown
that the singular CH contracts to meet the central $r=0$ {\it spacelike}
singularity. More recently, Burko \cite{Burko} studied the same 
model problem.
His work improve the numerical results
given in \cite{Brady}, namely, he found a good agreement between the
non-linear numerical results and the predictions of the perturbative approach.

Despite the important results achieved in these numerical 
investigations the mass-inflation scenario has {\it never}
been demonstrated {\it explicitly} in {\it collapsing} situation. It 
should be emphasized that these numerical works
begin on a {\it singular} Reissner-Nordstr\"om spacetime 
and the black-hole formation was {\it not} calculated there.
The main goal of this paper is to demonstrate {\it explicitly} that 
mass-inflation takes place during a dynamical charged gravitational collapse.

We consider the gravitational collapse of a self-gravitating {\it charged} 
scalar-field $\phi$.
While charged collapse is not expected in a realistic gravitational
collapse, it is generally accepted that the similarity between the
inner structure of a Reissner-Nordstr\"om black-hole and a Kerr
black-hole indicates that a charged collapse might be a simple
(spherical) toy model for the more realistic generic rotating collapse.
In previous papers we have investigated the {\it linear} \cite{HodPir1,HodPir2}
and {\it non}-linear \cite{HodPir3} evolution of a {\it charged}
scalar-field {\it outside} a charged black-hole.
The results given in these papers, and in particular the 
existence of oscillatory inverse {\it power-law}
charged tails along the black-hole outer horizon 
suggest the occurrence of mass-inflation 
along the CH of a {\it dynamically} formed charged
black-hole. Thus, this model is suitable to establish  
our main goal.

Our scheme is based on {\it double null} coordinates.
This allows us to begin with {\it regular}
initial spacetime (at approximately past null infinity),
calculate the {\it formation} of the black-hole's
event horizon,
and follow the evolution {\it inside} the
black-hole all the way to the central singularity, which 
is formed {\it during} the collapse. 
Thus, this numerical scheme makes it possible to test {\it directly} (and
for the first time) the conjecture that 
the mass-inflation scenario is an inevitable feature of a generic 
gravitational collapse.

The physical model is described by
the coupled Einstein-Maxwell-charged scalar equations.
We express the metric of a spherically symmetric spacetime in the
form \cite{Christodoulou}
\begin{equation}\label{Eq1}
ds^{2}=-\alpha(u,v)^{2}dudv+r(u,v)^{2}d\Omega ^{2}\  ,
\end{equation}
where $u$ is a retarded time null coordinate and $v$ is an
advanced time null coordinate.
We fix the axis $r=0$ to be along $u=v$.
The remaining coordinate freedom is the freedom
of the choice
of $v$ along some future null cone, i.e. along some fixed $u=const$ 
outgoing null ray. It should be noted that for $v \gg M$ our null 
ingoing coordinate $v$ is proportional to the Eddington-Finkelstein
null ingoing coordinate $v_{e}$.
Following Hamade and Stewart \cite{Hamade} we formulate the problem as
a system of {\it first}-order coupled PDEs. To do so we define 
auxiliary variables $d,f,g,s,x$ and $y$:
\begin{equation}\label{Eq2}
d={\alpha_{v} \over \alpha},\ \
f=r_{u},\ \ g=r_{v}, \ \ s=\sqrt{4\pi} \phi,\ \ x=s_{u},\ \ y=s_{v}\ .
\end{equation}
Using these variables one can generalize the neutral Hamade and
Stewart scheme \cite{Hamade}.
In terms of these new variables the Einstein equations expand to
\begin{equation}\label{Eq3}
E1 \equiv d_{u}-{{fg} \over {r^{2}}}-{\alpha^{2} \over {4r^{2}}}+
{{\alpha^{2}q^{2}} \over {2r^{4}}}+{1 \over 2}(xy^{*}+x^{*}y)+
{1 \over 2}iea(sy^{*}-s^{*}y)=0\  ,
\end{equation}
\begin{equation}\label{Eq4}
E2 \equiv rf_{v}+fg+{1 \over 4}\alpha^{2}-{{\alpha^{2}q^{2}} \over {4r^{2}}}=0\
\end{equation}
and
\begin{equation}\label{Eq5}
E3 \equiv g_{v}-2dg+ry^{*}y=0\  .
\end{equation}
The electromagnetic potential $a(u,v) \equiv A_{0}$ is given
by the Maxwell equations
\begin{equation}\label{Eq6}
M1 \equiv a_{v}-{{\alpha^{2}q} \over {2r^{2}}}=0\  ,
\end{equation}
where the charge $q(u,v)$ is given by
\begin{equation}\label{Eq7}
M2 \equiv q_{v}-ier^{2}(s^{*}y-sy^{*})=0\  .
\end{equation}
The Hawking mass function $m(u,v)$ is given by
\begin{equation}\label{Eq8}
m={r \over 2}(1+{q^2 \over r^2}+{4 \over \alpha^2}r_{u}r_{v})\  .
\end{equation}
Finally, the wave-equation for the charged scalar-field becomes
\begin{equation}\label{Eq9}
S1 \equiv ry_{u}+fy+gx+ieary+ieags+{{ie} \over {4r}} \alpha^{2} qs=0\  .
\end{equation}
The definition-equation (\ref{Eq2}) yields
\begin{equation}\label{Eq10}
D1 \equiv d-{\alpha_{v} \over \alpha}=0, \ \ \  D2 \equiv  g-r_{v}=0,
 \ \ \ D3 \equiv  y-s_{v}=0\  .
\end{equation}

The initial conditions include the specification of $y(0,v)$ and
$d(0,v)$ along an outgoing $u=0$ null ray. We assume $d(0,v)=0$, which
fixes the remaining freedom in the coordinates.
The boundary conditions on the axis $r=0$ ($u=v$) are $g=-f={1 \over
  2}\alpha$, $x=y$, $a=q=0$ and $\alpha_{r}=s_{r}=0$ (on axis).

The evolution of the quantities $d$ and $y$ are determined by
Eqs. E1 and S1, respectively.
We then integrate Eq. D1 outwards from the axis along an
outgoing ($u=const$) null ray to find $\alpha$.
Eqs. E3 and D2 are used to obtain $g$ and $r$.
Finally, we evaluate the quantities $s,q,a,f$ and $x$ by 
integrating Eqs. D3, M2, M1, E2 and S1, respectively ($x$ is 
evaluated from S1 using the relation $x_{v}=y_{u}$).  
The integration in the $u$-direction is carried out using a
fifth-order Runge-Kutta method, while the integrals in the $v$-direction
are discretized using the three-point Simpson method \cite{Press}.

\begin{figure}
\begin{center}
\includegraphics[width=12cm]{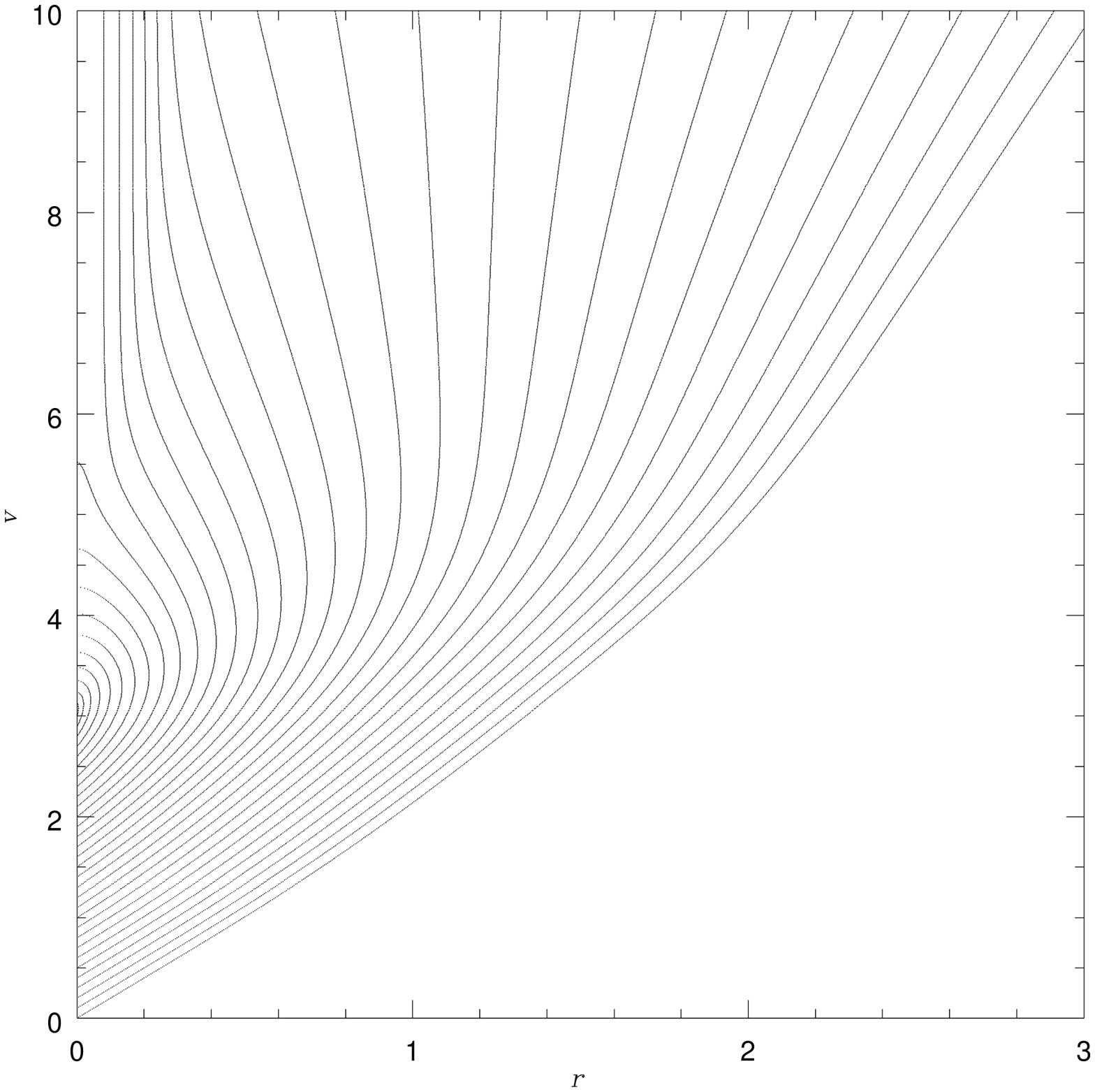}
\caption[null rays in the $rv$-plane]{\label{Fig1}
Null rays in the $rv$-plane. One can distinguish between three types of
outgoing null rays: The outer-most, which escape to infinity, the
inner-most which terminate at the singular section of the $r=0$
hypersurface and the intermediate outgoing null rays which approach a
($u$-dependent) finite radius, indicating the existence of a CH.
All the null rays originate from the {\it non-}singular axis $r=0$.}
\end{center}
\end{figure}

Figure \ref{Fig1} displays the radius $r(u,v)$ as a function of the
ingoing null coordinate $v$ along a sequence of outgoing ($u=const$)
null rays.  All the outgoing null rays originate from the {\it
  non}-singular axis $r=0$, i.e. we start with a {\it regular}
spacetime (this situation is in contrast with previous numerical
works, where the evolution begins on a Reissner-Nordstr\"om
spacetime).  One can distinguish between {\it three} types of outgoing
null rays in the $rv$ plane: (i) The outer-most (small-$u$) rays,
which {\it escape} to infinity.  (ii) The intermediate outgoing null
rays approach a fixed radius $r_{CH}(u)$ at late-times $v \to \infty$.
This indicates the existence of a CH in these spacetimes.  (iii) The
inner-most (large-$u$) rays, which originate on the {\it non}-singular
axis $r=0$ and terminate at the {\it singular} section of the $r=0$
hypersurface.  These outgoing rays reach the $r=0$ singularity in a
{\it finite} $v$, {\it without} intersecting the CH.  This situation
is in contrast with the Reissner-Nordstr\"om spacetime, in which it is
well-known that all the outgoing null rays which originate inside the
black-hole intersect the CH.  One should also note that in contrast
with the Reissner-Nordstr\"om spacetime, where the CH is a {\it
  stationary} null hypersurface, here $r_{CH}(u)$ depends on the
outgoing null coordinate $u$, i.e. the CH {\it contracts}
\cite{Brady}. This dramatic difference in the causal structure of the
present {\it collapsing} spacetime compared with the
Reissner-Nordstr\"om spacetime is attributed to the outgoing flux of
energy-momentum carried by the charged scalar-field.

\begin{figure}
\begin{center}
\includegraphics[width=12cm]{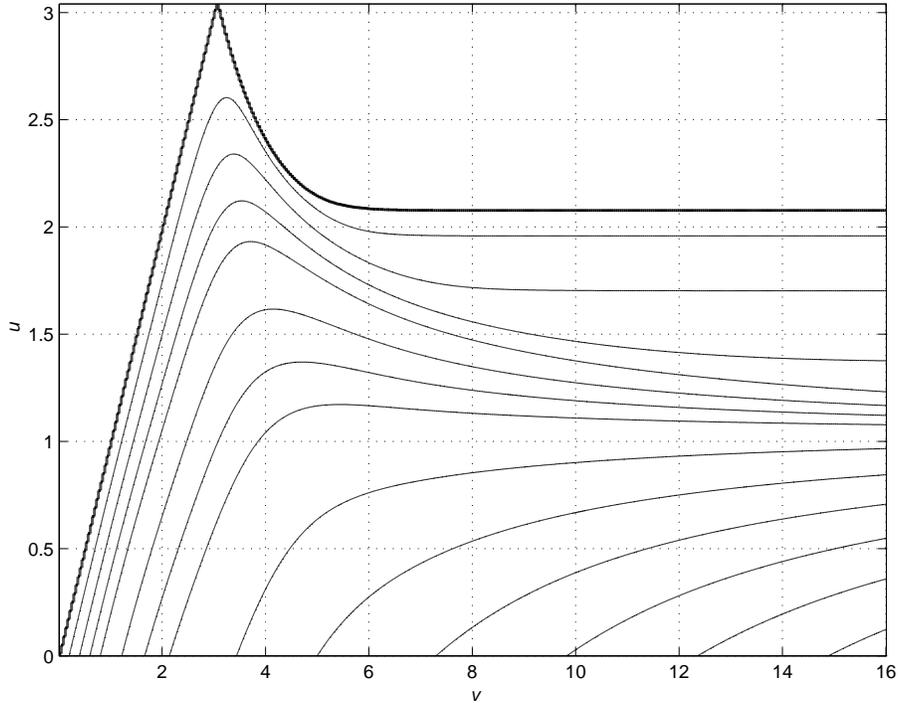}
\caption[Contour lines of the coordinate $r$ in the $vu$-plane]{\label{Fig2}
Contour lines of the coordinate $r$ in the $vu$-plane.
The $r=0$ contour line is indicated by a thicker curve. Its left
section ($u=v$) represents the {\it non-}singular axis, while its
right section corresponds the the central {\it spacelike}
singularity. The apparent horizon is indicated by the vanishing of
$r_{v}$. The (singular) CH (a null hypersurface, located 
at $v \to \infty$, and indicated by the
approach of outgoing null-rays to {\it finite} values of $r$) 
{\it contracts} to meet the {\it central} spacelike singularity (in a
{\it finite} proper time).}
\end{center}
\end{figure}

To understand better the causal structure of our dynamical spacetime
we display in Fig. \ref{Fig2}. the countor lines of $r(u,v)$ in the
$vu$-plane. The outer most contour line corresponds to $r=0$, where
its left section (a straight line $u=v$) is the {\it non}-singular
axis, and its right section corresponds to the central singularity at
$r=0$.  It should be emphasized that this central singularity forms
{\it during} the gravitational collapse.  The singularity at the $r=0$
hypersurface is clearly a {\it spacelike} one for $r_{v}$ is {\it
  negative} along this section.  The vanishing of $r_{v}$ indicates
the existence of an apparent horizon (which is first formed at $u
\approx 1$ for this specific solution).  The CH itself is a {\it null}
hypersurface which is located at $v \to \infty$. Its existence is
indicated by the fact that the intermediate outgoing null rays (in the
range $1 \lesssim u \lesssim 2.1$ for this specific solution)
terminate at a finite ($u$-dependent) radius $r_{CH}(u)$.  The
singular CH contracts to meet the central ($r=0$) spacelike
singularity (along the $u \simeq 2.1$ outgoing null ray).  Thus, the
{\it null} CH singularity is a precursor of the final {\it spacelike}
singularity along the $r=0$ hypersurface \cite{Brady}.

\begin{figure}
\begin{center}
\includegraphics[width=12cm]{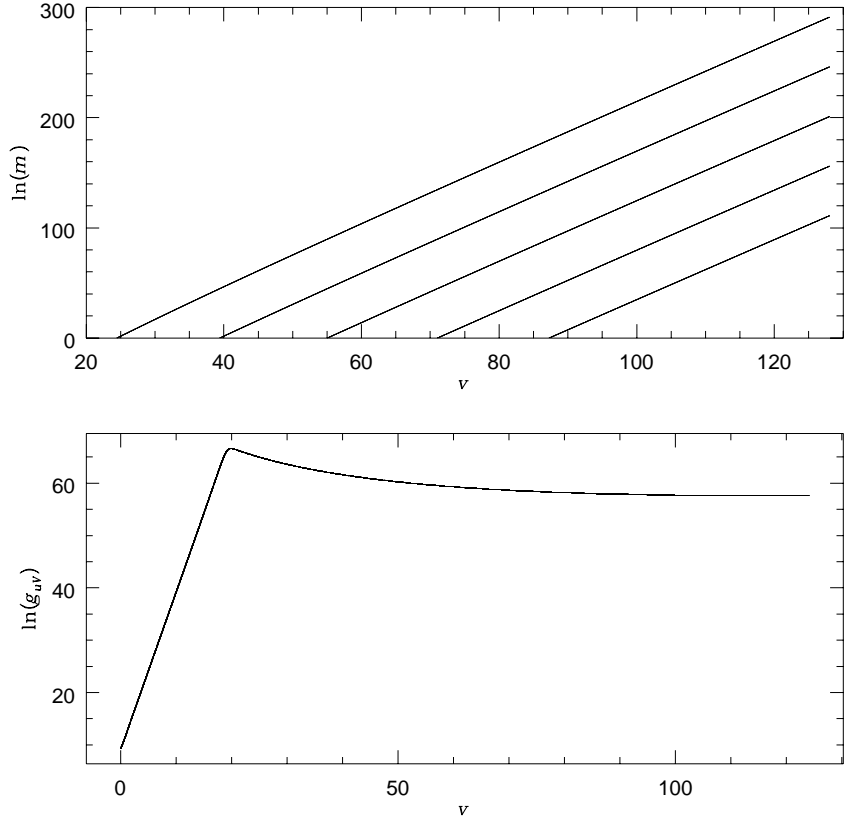}
\caption[mass-inflation]{\label{Fig3}
The CH singularity.
The top panel displays ln($m$) vs. advanced time $v$, 
along a sequence of outgoing null
rays. The exponential growth of the mass-function demonstrates 
the appearance of the mass-inflation scenario \cite{Poisson}.
The bottom panel displays
the metric function $g_{uV}$ along an outgoing null ray.
The finite value approached by the metric functions is in agreement
with the simplified model of Ori \cite{Ori1,Ori2}, and demonstrate 
the {\it weakness} of the null mass-inflation singularity.}
\end{center}
\end{figure}

The behaviour of the mass function $m(u,v)$ along the outgoing null
rays is displayed in the top panel of Fig. \ref{Fig3}.  This figure
establish {\it explicitly} the exponential divergence of the mass
function (and curvature) along the CH in a {\it dynamically}
collapsing spacetime. To our knowledge, this is the {\it first}
explicit demonstration of the mass-inflation scenario in a {\it
  collapsing} situation starting from a {\it regular} spacetime.  The
mass function increases not only along the outgoing ($u$=const) null
rays (as $v$ increases) but also along ingoing ($v$=const) null rays
(as $u$ increases).

The {\it weakness} of the null mass-inflation singularity was first
predicted by Ori \cite{Ori1,Ori2}. It is demonstrated in the bottom
panel of Fig. \ref{Fig3}. This figure displays the metric function
$g_{uV}$ along an outgoing null ray, where $V$ is a Kruskal-like
ingoing null coordinate. Clearly, $g_{uV}$ approaches a {\it finite}
value as the CH is approached ($V \to 0$). This confirms the
analytical analysis of Ori, according to which a suitable coordinate
transformation can produce a {\it non-}singular metric.

\begin{figure}
\begin{center}
\includegraphics[width=12cm]{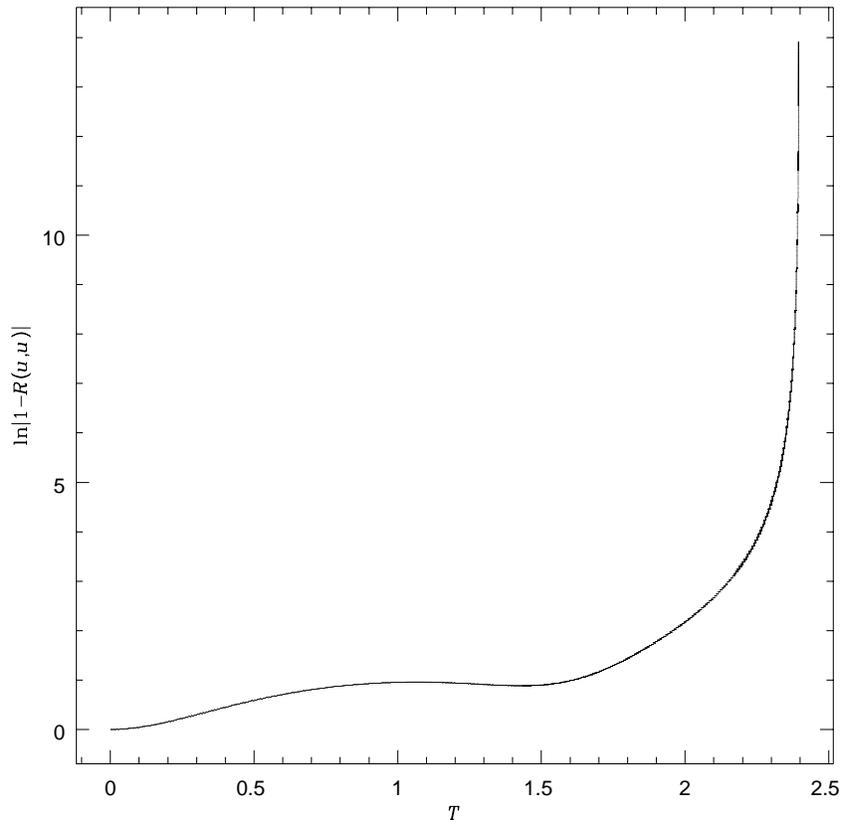}
\caption[Ricci curvature scalar on axis]{\label{Fig4}
The Ricci curvature scalar $R(u,u)$ as a function of 
the proper time on axis.
The divergence of the curvature on the axis indicates that 
the {\it initially regular} axis
is replaced in a {\it finite} proper time by a 
central (spacelike) {\it singularity}.}
\end{center}
\end{figure}

Figure \ref{Fig4} displays the Ricci curvature scalar 
(on the axis):
\begin{equation}\label{Eq11}
R(u,v)=-{8 \over {\alpha^{2}}} \left [ Re(y^{*}x)+eaIm(s^{*}y) \right
]\  ,  
\end{equation}
as a function of $T$, the proper time on axis \cite{Hamade}:
\begin{equation}\label{Eq12}
T(u)= \int_{0}^{u} {\alpha(w,w)dw}\  .
\end{equation}
The curvature on the axis diverges in a {\it finite} proper time.
Thus, the {\it initially regular} axis is replaced by a (spacelike)
{\it singularity} along the $r=0$ hypersurface.

In summary, we have studied the gravitational collapse of a
self-gravitating {\it charged} scalar-field. We calculate the
formation of an apparent horizon, followed by a {\it weak}, null, {\it
  mass-inflation} singularity along the {\it contracting} CH, which
precedes a {\it strong, spacelike} singularity along the $r=0$
hypersurface.  Our results give a first {\it explicit} confirmation of
the mass-inflation scenario in a dynamical collapse that begins with
{\it regular} initial conditions.

\bigskip
\noindent
{\bf ACKNOWLEDGMENTS}
\bigskip

This research was supported by a grant from the Israel Science Foundation.
TP thanks W. Israel for helpful discussions.

\newpage
\newpage

\newpage

\end{document}